\documentclass[aps,pre,twocolumn,nofootinbib,floatfix]{revtex4}
\usepackage{amsmath}
\usepackage{epsfig}

\input{tcilatex}

\begin{document}

\title{Superradiant and Aharonov-Bohm effect for the quantum ring exciton }
\author{Y. N. Chen and D. S. Chuu}
\affiliation{Department of Electrophysics, National, Chiao-Tung University, Hsinchu 300,
Taiwan}
\date{\today }

\begin{abstract}
The Aharonov-Bohm and superradiant effect on the redaitive decay rate of an
exciton in a quantum ring is studied. With the increasing of ring radius,
the exciton decay rate is enhanced by superradiance, while the amplitude of
AB oscillation is decreased. The competition between these two effects is
shown explicitly and may be observable in time-resolved exeriments.

PACS: 42.50.Fx, 32.70.Jz, 71.35.-y, 71.45.-d
\end{abstract}

\maketitle

\address{$^{1}$Department of Electrophysics, National Chiao Tung University,
Hsinchu 30050, Taiwan\\
$^{2}$Department of Physics, UMIST, P.O. Box 88, Manchester, M60 1QD, U.K.}

\address{$^{1}$Department of Electrophysics, National Chiao Tung University,
Hsinchu 30050, Taiwan} 
\address{$^{2}$Department of Physics, UMIST, P.O. Box 88, Manchester, M60
1QD, U.K.}

\address{Department of Electrophysics, National Chiao Tung University,
Hsinchu 300, Taiwan}





With the advances of modern fabrication technologies, it has become possible
to fabricate the ring-shaped dots of InAs in GaAs \cite{1}. In the
circumstances of Aharonov-Bohm (AB) effect, one of the important features is
the periodic dependence of interference patterns on magnetic flux $\Phi $%
\cite{2}. Most of the measurements, however, are available only from the
transport experiments on metallic rings in the mesoscopic regime \cite{3}.
Very recently, optical detection of the AB effect on an exciton in a single
quantum ring has become possible \cite{4}. This makes it more interesting to
study the optical properties of the quantum ring exciton.

On the other hand, the electron-hole pair is naturally a candidate for
examining the spontaneous emission. However, as it was well known, the
excitons in a three dimensional system will couple with photons to form
polaritons--the eigenstate of the combined system consisting of the crystal
and the radiation field which does not decay radiatively\cite{5}. Thus, in a
bulk crystal, the exciton can only decay via impurity, phonon scatterings,
or boundary effects. The exciton can render radiative decay in lower
dimensional systems such as quantum wells, quantum wires, or quantum dots as
a result of broken symmetry. The decay rate of the exciton is superradiant
enhanced by a factor of $\lambda /d$ in a 1D system\cite{6} and ($\lambda
/d)^{2}$ for 2D exciton-polariton\cite{7}, where $\lambda $ is the wave
length of emitted photon and $d$ is the lattice constant of the 1D system or
the thin film. In the past decades, the superradiance of excitons in these
quantum structures have been investigated intensively\cite{8}.

Although many investigations have been focused on superradiance of the
quantum confined excitons, the coherent radiation together with the AB
effect for an exciton in the ring geometry has received little attention so
far. In this paper, we investigate the decay properties of a neutral exciton
in the one-dimensional quantum ring. It is found that there is a competition
between the superradiant and AB effect for the exciton decay rate.

Consider first an exciton in a quantum ring with radius $\rho \sim Nd/2\pi $%
, where $d$ is the lattice spacing and $N$ is the number of the lattice
points. In our model, the circular ring is joined by the $N$ lattice points,
and we also assume the effective mass approximation is valid in the
circumference direction. The validity of these assumptions will be discussed
later. Therefore, the state of the exciton can be specified as $\left| \nu
,n,m\right\rangle $, where $\nu $ is the exciton wave number. $n$ and $m$
are quantum numbers for internal structure of the exciton, and will be
specified later. Here, $\nu $ takes the value of an integer. The matter
Hamiltonian can be written as

\begin{equation}
H_{ex}=\sum_{\nu nm}E_{\nu nm}c_{\nu nm}^{\dagger }c_{\nu nm},
\end{equation}
where $c_{\nu nm}^{\dagger }$ and $c_{\nu nm}$ are the creation and
destruction operators of the exciton, respectively. The Hamiltonian of free
photon is

\begin{equation}
H_{ph}=\sum_{\mathbf{q}^{\prime }k_{z}^{\prime }\lambda }\hbar c(q^{\prime
2}+k_{z}^{\prime 2})^{1/2}b_{\mathbf{q}^{\prime }k_{z}^{\prime }\lambda
}^{\dagger }b_{\mathbf{q}^{\prime }k_{z}^{\prime }\lambda },
\end{equation}
where $b_{\mathbf{q}^{\prime }k_{z}^{\prime }\lambda }^{\dagger }$ and $b_{%
\mathbf{q}^{\prime }k_{z}^{\prime }\lambda }$ are the creation and
destruction operators of the photon, respectively. The wave vector $\mathbf{k%
}^{\prime }$\hspace{0.06in}of the photon were separated into two parts: $%
k_{z}^{\prime }$ is the perpendicular component of $\mathbf{k}^{\prime }$ on
the ring plane such that $k^{\prime 2}=q^{\prime 2}+k_{z}^{\prime 2}$.

The interaction between the exciton and the photon can be expressed as

\begin{eqnarray}
H^{\prime } &=&\sum_{i}\sum_{\mathbf{q}^{\prime }k_{z}^{\prime }\lambda }%
\frac{e}{mc}\sqrt{\frac{2\pi \hbar c}{(q^{\prime 2}+k_{z}^{\prime 2})^{1/2}v}%
} \\
&&\times \lbrack b_{\mathbf{q}^{\prime }k_{z}^{\prime }\lambda }^{\dagger
}H_{\nu }^{(1)}(q^{\prime }\rho )\exp (i\nu ^{\prime }\varphi _{i})+\mathbf{%
h.c}.](\mathbf{\epsilon }_{\mathbf{q}^{\prime }k_{z}^{\prime }\lambda }\cdot 
\mathbf{p}_{i}),  \notag
\end{eqnarray}%
where $H_{\nu }^{(1)}$ is the Hankel function, $(\mathbf{\rho },\varphi
_{i}) $ is the position of the electron $i$ in the ring, $\mathbf{p}_{i}$ is
the corresponding momentum of the electron $i$\ operator, and $\mathbf{%
\epsilon }_{\mathbf{q}^{\prime }k_{z}^{\prime }\lambda }$ is the
polarization vector of the photon. The using of Hankel function in Eq. (3)
means the wave which generated by the recombination of the exciton moves
outward to infinity \cite{9}. For large radius, the Hankel function behaves
like $e^{iq^{\prime }\rho }.$

The exciton state in a quantum ring can be expressed as

\begin{equation}
\left| \nu ,n,m\right\rangle =\sum_{\varphi _{e},\varphi _{h}}U_{\nu
,n,m}^{*}(\varphi _{e},\varphi _{h})\left| c,\varphi _{e}+\varphi
_{h};v,\varphi _{h}\right\rangle ,
\end{equation}
and the interaction matrix elements can be written as

\begin{eqnarray}
\left\langle \nu ,n,m\right| H^{\prime }\left| G\right\rangle
&=&\sum_{\varphi _{e},\varphi _{h}}\left\langle c,\varphi _{e}+\varphi
_{h};v,\varphi _{h}\right|  \notag \\
&&\times U_{\nu ,n,m}^{\ast }(\varphi _{e},\varphi _{h})H^{\prime }\left|
G\right\rangle ,
\end{eqnarray}%
in which the excited state $\left| c,\varphi _{e}+\varphi _{h};v,\varphi
_{h}\right\rangle $ is defined as

\begin{equation}
\left| c,\varphi _{e}+\varphi _{h};v,\varphi _{h}\right\rangle =a_{c,\varphi
_{e}+\varphi _{h}}^{\dagger }a_{v,\varphi _{h}}\left| G\right\rangle ,
\end{equation}
where $a_{c,\varphi _{e}+\varphi _{h}}^{\dagger }$ ($a_{v,\varphi _{h}}$) is
the creation (destruction) operator for an electron (hole) in the conduction
(valence) band at site $\varphi _{e}+\varphi _{h}(\varphi _{h})$. The
expansion coefficient $U_{\nu ,n,m}^{*}(\varphi _{e},\varphi _{h})$ is the
exciton wave function in the quantum ring:

\begin{equation}
U_{\nu ,n,m}^{\ast }(\varphi _{e},\varphi _{h})=\frac{1}{\sqrt{N}}\exp (i\nu
r_{c})F_{nm}(\varphi _{e}),
\end{equation}%
where the coefficient $1/\sqrt{N}$ is for the normalization of the state $%
\left| \nu ,n,m\right\rangle $, and $r_{c}=\frac{m_{e}^{\ast }(\varphi
_{e}+\varphi _{h})+m_{h}^{\ast }\varphi _{h}}{m_{e}^{\ast }+m_{h}^{\ast }}$
is the center of mass of the exciton. Here, $m_{e}^{\ast }$ and $m_{h}^{\ast
}$ are, respectively, the effective masses of the electron and the hole. $%
F_{nm}(\varphi _{e})$ is the hydrogenic wavefunction in the ring and will be
calculated later.

After summing over $\varphi _{h}$, we have

\begin{eqnarray}
\left\langle \nu ,n,m\right| H^{\prime }\left| G\right\rangle &=&\sum_{%
\mathbf{q}^{\prime }k_{z}^{\prime }\lambda }\frac{e}{mc}\sqrt{\frac{2\pi
\hbar c}{(q^{\prime 2}+k_{z}^{\prime 2})^{1/2}v}} \\
&&\times \lbrack b_{\mathbf{q}^{\prime }k_{z}^{\prime }\lambda }(\mathbf{%
\epsilon }_{\mathbf{q}^{\prime }k_{z}^{\prime }\lambda }\cdot \mathbf{A}%
_{\nu nm})H_{\nu }^{(1)}+\mathbf{h.c}.],  \notag
\end{eqnarray}%
where

\begin{eqnarray}
\mathbf{A}_{\nu nm} &=&\sqrt{N}\sum_{\varphi _{e}}F_{nm}(\varphi _{e})\int d%
\mathbf{\varphi }w_{c}(\varphi -\varphi _{e})  \notag \\
&&\times \exp (i\nu (\varphi -\frac{m_{e}^{\ast }\varphi _{e}}{m_{e}^{\ast
}+m_{h}^{\ast }}))(-i\hbar \frac{\partial }{\partial \varphi })w_{v}(\varphi
).
\end{eqnarray}%
Here, $w_{c}(\varphi )$ and $w_{v}(\varphi )$ are, respectively, the Wannier
functions for the conduction band and the valence band.

The essential quantity involved is the matrix element of $H^{\prime }$
between the ground state $\left| G\right\rangle $ and the exciton state $%
\left| \nu ,n,m\right\rangle $. Hence the interaction between the exciton
and the photon (in the resonance approximation) can be written in the form

\begin{equation}
H^{\prime }=\sum_{k_{z}^{\prime }nm}\sum_{\mathbf{q}^{\prime }\lambda }D_{%
\mathbf{q}^{\prime }k_{z}^{\prime }\nu nm}b_{k_{z}^{\prime }\mathbf{q}%
^{\prime }\lambda }c_{\nu nm}^{\dagger }+\mathbf{h.c.,}
\end{equation}
where

\begin{equation}
D_{\mathbf{q}^{\prime }k_{z}^{\prime }\nu nm}=H_{\nu }^{(1)}(q^{\prime }\rho
)\frac{e}{mc}\sqrt{\frac{2\pi \hbar c}{(q^{\prime 2}+k_{z}^{\prime 2})^{1/2}v%
}}\mathbf{\epsilon }_{\mathbf{q}^{\prime }k_{z}^{\prime }\lambda }\cdot 
\mathbf{A}_{\nu nm}.
\end{equation}

By the method of Heitler and Ma in the resonance approximation, the decay
rate of the exciton can be expressed as

\begin{equation}
\gamma _{_{\nu nm}}=2\pi \sum_{\mathbf{q}^{\prime }k_{z}^{\prime }\lambda
}\left| D_{\mathbf{q}^{\prime }k_{z}^{\prime }\nu nm}\right| ^{2}\delta
(\omega _{\mathbf{q}^{\prime }k_{z}^{\prime }\nu nm}),
\end{equation}
where $\omega _{\mathbf{q}^{\prime }k_{z}^{\prime }\nu nm}=E_{\nu nm}/\hbar
-c\sqrt{q^{\prime 2}+k_{z}^{\prime 2}}.$

The exciton decay rate in the optical region can be calculated
straightforwardly and is given by

\begin{eqnarray}
\gamma _{\nu nm} &=&\frac{e^{2}\hbar }{m^{2}c}\frac{\rho }{d}\int \left|
H_{\nu }^{(1)}(q^{\prime }\rho )\right| ^{2}q^{\prime } \\
&&\times \int \frac{\delta (\omega _{\mathbf{q}^{\prime }k_{z}^{\prime }\nu
nm})}{\sqrt{k_{z}^{\prime 2}+q^{\prime 2}}}\left| \mathbf{\epsilon }_{%
\mathbf{q}^{\prime }k_{z}^{\prime }\lambda }\cdot \mathbf{\chi }_{\nu
nm}\right| ^{2}dk_{z}^{\prime }dq^{\prime },  \notag
\end{eqnarray}%
where

\begin{equation}
\mathbf{\chi }_{\nu nm}=\sum_{\varphi _{e}}F_{nm}^{*}(\varphi _{e})\int d%
\mathbf{\varphi }w_{c}^{*}(\varphi -\varphi _{e})(-i\hbar \frac{\partial }{%
\partial \varphi })w_{v}(\mathbf{\varphi }).
\end{equation}

Here, $\mathbf{\chi }_{\nu nm}^{\ast }$ represents the effective dipole
matrix element for an electron jumping from the excited Wannier state in the
conduction band back to the hole state in the valence band. As one can see
from Eq. (13), the decay rate $\gamma _{\nu nm}$ is proportional to $\rho /d$%
. This is just the superradiant factor implying the coherent contributions.
Furthermore, the asymptotic limit of $\gamma _{\nu nm}$ ($\rho \rightarrow
\infty $) recovers the exciton decay rate in one-dimensional quantum wire: $%
R_{1d}=\frac{3\pi }{2k_{0}d}\gamma _{0}$, where $k_{0}=2\pi /\lambda $ $\ $%
and $\gamma _{0}$ is the decay rate of an isolated two level atom.

Now let us consider the AB effect for a superradiant exciton. For the
one-dimensional quantum ring, the exciton wavefunction can be solved by R%
\"{o}mer and M. E. Raikh's approach\cite{10}. The wavefunction in the ground
state$(n=m=0)$ can be expressed as,

\begin{equation}
F_{00}(0)=[V_{0}^{2}\sum_{N^{\prime }=1}^{\infty }\frac{1}{(E_{N^{\prime
}}^{(e)}+E_{-N^{\prime }}^{(h)}-\bigtriangleup _{0}^{0})^{2}}]^{-1},
\end{equation}
where $E_{N^{\prime }}^{(e)}=\frac{\hbar ^{2}}{2m_{e}\rho ^{2}}(N^{\prime }-%
\frac{\Phi }{\Phi _{0}})$ and $E_{-N^{\prime }}^{(h)}=\frac{\hbar ^{2}}{%
2m_{e}\rho ^{2}}(N^{\prime }+\frac{\Phi }{\Phi _{0}})$ with the universal
flux $\Phi _{0}=hc/e$. The constant $V_{0}<0$ is defined as

\begin{equation}
V_{0}=\frac{1}{2\pi }\int d\varphi V[R(\varphi )].
\end{equation}
And the exciton energy $\bigtriangleup _{0}^{0}$ takes the form

\begin{equation}
(\frac{\bigtriangleup _{0}^{0}}{\varepsilon _{0}})^{1/2}=-(\frac{\pi V_{0}}{%
\varepsilon _{0}})\frac{\sin (2\pi (\frac{\bigtriangleup _{0}^{0}}{%
\varepsilon _{0}})^{1/2})}{\cos (2\pi (\frac{\bigtriangleup _{0}^{0}}{%
\varepsilon _{0}})^{1/2})-\cos (2\pi (\frac{\Phi }{\Phi _{0}}))},
\end{equation}%
where $\varepsilon _{0}=\frac{\hbar ^{2}}{2\rho ^{2}}(\frac{1}{m_{e}}+\frac{1%
}{m_{h}})$.

In the limit of large radius, the corresponding ground state energy is

\begin{equation}
\bigtriangleup _{0}^{0}=-\frac{\pi ^{2}V_{0}^{2}}{\varepsilon _{0}}[1+4\cos (%
\frac{2\pi \Phi }{\Phi _{0}})\exp (-\frac{2\pi ^{2}\left| V_{0}\right| }{%
\varepsilon _{0}})].
\end{equation}%
One should note $V[R(\varphi )]$ is not specified since it describes the
interaction between the electron and hole in a realistic quantum ring.
However, $V_{0}$ can still be extracted from Eq. (18) in large radius limit,
i.e. applying real experimental data of a \emph{quantum wire} exciton energy 
$\bigtriangleup _{0}^{0}$. Besides, the exponential factor in Eq. (18) can
be represented by $exp(-2\pi \rho /l$) \cite{10}, where $l$ is the decay
length of the wave function of the internal motion of electron and hole.
Thus, the magnitude of the AB effect in the limit of large radius represents
the amplitude for bound electron and hole to tunnel in the opposite
directions and meet each other on the opposite side of the ring. 
\begin{figure}[th]
\includegraphics[width=8cm]{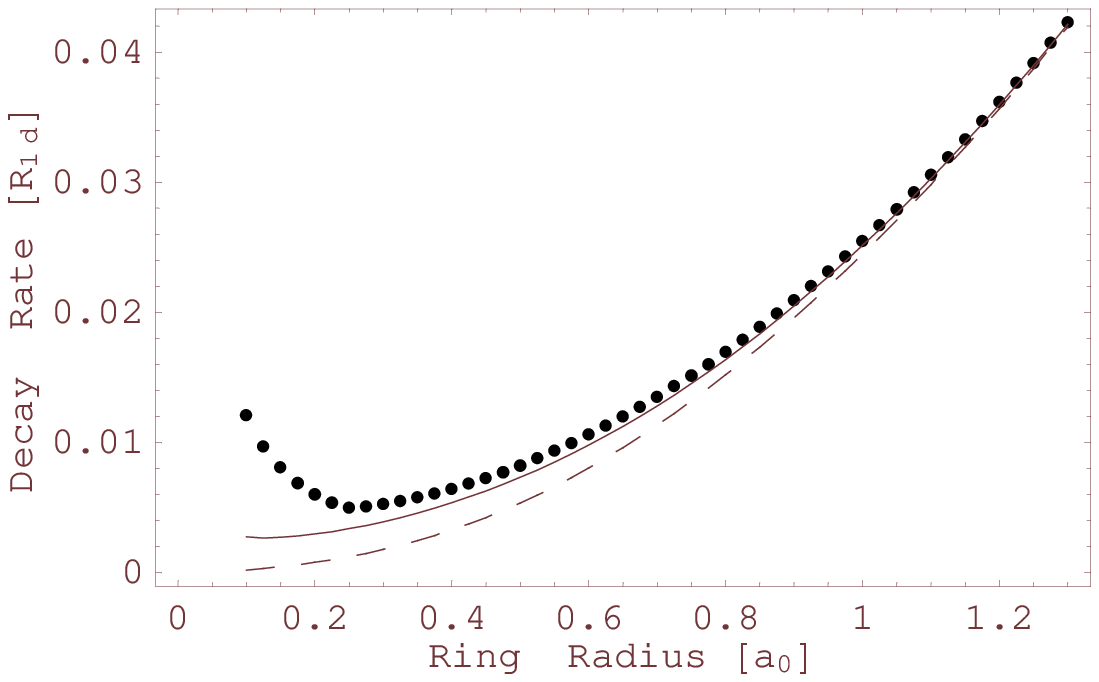}
\caption{Effect of Aharonov-Bohm on the radiative decay of a quantum ring
exciton. The dashed(-- --), solid, and dotted( $\cdot $) curves correspond
to $\Phi =0\Phi _{0},$ $0.25\Phi _{0},$ and $0.5\Phi _{0},$ respectively. In
small radius limit, $F_{nm}^{\ast }$ depends strongly on radius $\protect%
\rho $, and its influence on the decay rate is evident. The vertical and
horizontal units here are $\frac{3\protect\pi }{2k_{0}d}\protect\gamma _{0}$
and ring radius (in units of $a_{0}$), respectively. }
\end{figure}

The dipole matrix element $\mathbf{\chi }_{\nu nm}$ in Eq. (14) corresponds
to an average of dipole transitions between different sites, weighted by the
exciton wave function $F_{nm}^{\ast }(\varphi _{e}).$ The sum in Eq. (14)
contains a $\varphi _{e}\rightarrow 0$ term in which the electron and the
hole are at the same site. If the corresponding integral does not vanish,
this term dominates the sum. The effective dipole transition matrix element
becomes

\begin{equation}
\mathbf{\chi }_{\nu nm}\sim F_{nm}^{\ast }(0)\int d\mathbf{\varphi }%
w_{c}^{\ast }(\varphi )(-i\hbar \mathbf{\nabla })w_{v}(\mathbf{\varphi }%
)=F_{nm}^{\ast }(0)\chi _{s},
\end{equation}%
where $\chi _{s}$ is essentially the dipole matrix element between the
atomic states at the same site. Combing Eqs. (13), (15), (17), and (19), one
can obtain the AB effect for a superradiant exciton. In Fig. 1 three curves
of different flux $\Phi $ are presented as a function of radius $\rho $. To
plot the figure, we have assumed the wavelength of the emitted photon $%
\lambda =8000\overset{\circ }{A}$ and lattice spacing $d=5\overset{\circ }{A}
$. The dashed, solid, and dotted curves represent the cases of $\Phi =0\Phi
_{0},$ $0.25\Phi _{0},$ and $0.5\Phi _{0},$ respectively. As can be seen, AB
effect becomes important in small radius limit. For $\Phi =0.5\Phi _{0}$,
the decay rate decreases with the decreasing of ring radius but reaches the
minimum point as $\rho $ is about 0.25$a_{0}$ (where $a_{0}=100\overset{%
\circ }{A}$ is the effective Bohr radius we assumed in 1D limit). This is
because the probability, for electron and hole to meet each other on the
opposite side of the ring, increases with the decreasing of ring radius,
while the coherent effect (superradiance) decreases with the decreasing of
the radius. As a result, there is a competition between these two effects.
One also notes the AB oscillation is not of constant amplitude. In Fig. 2,
relative decay rates [$\gamma _{_{\nu nm}}(\Phi )-\gamma _{_{\nu nm}}(\Phi
=0)$] as a function of magnetic flux $\Phi $ are plotted. The solid and
dashed lines represent the cases of $\rho =1$ $a_{0}$ and $\rho =0.5$ $a_{0}$
, respectively. The larger the radius, the smaller the AB oscillation
amplitude. As expected, the superradiant decay rate is most enhanced for $%
\Phi =0.5\Phi _{0}$, and the oscillation period is equal to $\Phi _{0}=hc/e$%
. 
\begin{figure}[th]
\includegraphics[width=8cm]{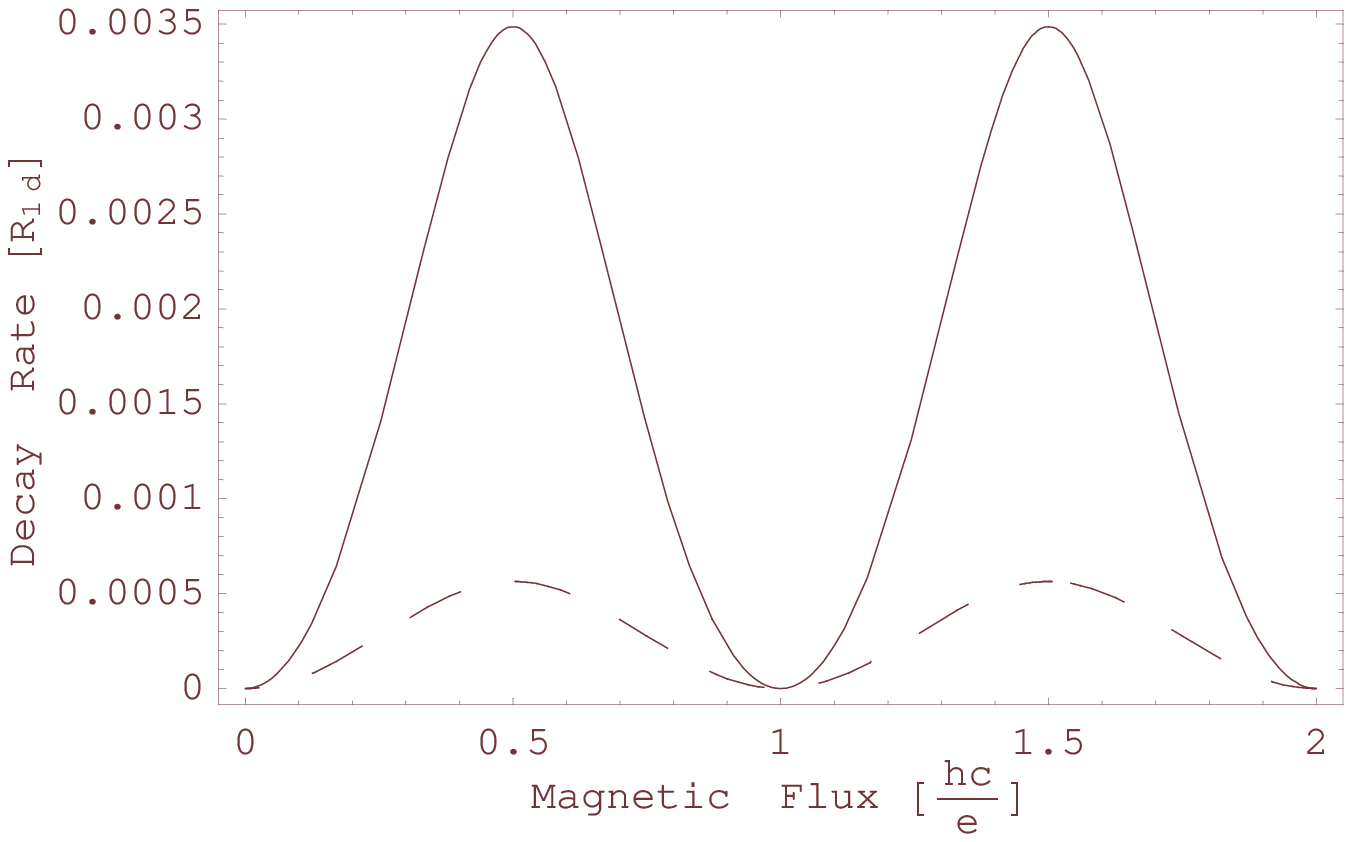}
\caption{Dependence of relative decay rate [$\protect\gamma _{_{\protect\nu %
nm}}(\Phi )-\protect\gamma _{_{\protect\nu nm}}(\Phi =0)$] on the magnetic
flux. The dashed and solid curves correspond to $\protect\rho =0.5$ $a_{0}$
and $\protect\rho =1$ $a_{0},$ respectively. The vertical and horizontal
units are $\frac{3\protect\pi }{2k_{0}d}\protect\gamma _{0}$ and universal
flux quantum $\Phi _{0}=hc/e,$ respectively.}
\end{figure}

Although present model considers the ideal one-dimensional quantum ring, the
physics discussed above can be applied to the realistic quantum ring with
finite width. The modified quantity is the exciton wavefunction $F_{nm},$
which only changes the amplitude of AB oscillation. In addition, the
coherent radiation from the lattice points within a wavelength still holds
as long as the angular momentum is preserved, i.e. not broken by impurities
or phonons. This means a high quality quantum is required to observe the
mentioned effects.

In summary, we have calculated the superradiant decay rate of an exciton in
a quantum ring. Flux dependent oscillation of the superradiant exciton is
shown explicitly. With the decreasing of ring radius, there is a competition
between the superradiant and AB effects. The distinguishing features are
pointed out and may be observed in a suitably designed experiment.

This work is supported partially by the National Science Council, Taiwan
under the grant number NSC 92-2120-M-009-010.

\end{document}